\documentclass[twoside,final]{pro12}

\usepackage[ansinew]{inputenc}
\usepackage{amsmath}
\usepackage{graphicx}
\usepackage{array}
\usepackage{SIunits}
\usepackage{lscape}
\usepackage{multirow}
\usepackage{float}
\usepackage{natbib}
\usepackage{hyperref}
\usepackage{xcolor}

\newenvironment{correction}{}

\head{J.-M. Grie{\ss}meier et al.} {On the required \correction{mass} for exoplanetary radio emission}	

\begin{document}

\title{On the required \correction{mass} for an exoplanet to emit radio waves}	

\author{
Jean-Mathias Grie{\ss}meier
	\adress{Laboratoire de Physique et Chimie de l'Environnement et de l'Espace (LPC2E) Universit\'{e} d'Orl\'{e}ans/CNRS, Orl\'{e}ans, France}$\,\,$, 
N. V. Erkaev
	\adress{The Applied Mechanics Department, Siberian Federal University, 660074 Krasnoyarsk, Russian Federation}
	$^{,}$\adress{Institute of Computational Modelling, Siberian Branch of the Russian Academy of Sciences, 660036 Krasnoyarsk, Russian Federation}
	$^{,}$\adress{Institute of Laser Physics, Siberian Branch of the Russian Academy of Sciences, 630090 Novosibirsk, Russian Federation}$\,\,$, 
C. Weber
	\adress{Space Research Institute, Austrian Academy of Sciences, Schmiedlstr 6, A-8042 Graz, Austria}$\,\,$, \\
H. Lammer $^4$, 
V. A. Ivanov $^{1,3}$, 
P. Odert
	\adress{Institute of Physics/IGAM, University of Graz, Universit\"{a}tsplatz 5, A-8010 Graz, Austria}
}

\maketitle

\begin{abstract}
The detection of radio emission from an exoplanet would constitute the 
best way to determine its magnetic field. Indeed, the presence of a 
planetary magnetic field is a necessary condition for radio emission 
via the Cyclotron Maser Instability. The presence of a magnetic field is, 
however, not sufficient. At the emission site, the local cyclotron 
frequency has to be sufficiently high compared to the local plasma 
frequency. 
\correction{As strong stellar insolation on a low-mass planet 
can lead to an extended planetary 
atmosphere, the magnetospheric plasma frequency depends
on the planetary 
mass, its orbital distance, and its host star.} 
\correction{We show that an extended planetary atmosphere can 
quench the radio emission. 
This seems to be true, in particular, for an important fraction of the 
planets less massive than 
approximately two Jupiter masses and with orbital distances below $\sim$0.2 AU.
Most of the best candidates suggested by radio scaling laws lie in this parameter
space. Taking this effect quenching into account will have important implications
for the target selection of observation 
campaigns. At the same time, this effect will have 
consequences for the interpretation of observational data.}
\end{abstract}

\section{Introduction}

In the solar system, planets with a magnetic field emit low-frequency, 
coherent, polarized radio emission via the so-called 
Cyclotron Maser Instability
\citep[CMI, see e.g.][]{Zarka98,Farrell99,Ergun00,Treumann06}. 
%
For exoplanets with a sufficiently strong magnetic field, the same type 
of emission is expected. 
Such exoplanetary radio emission is expected to be detecable with the 
latest generation of radio telescopes \citep[see, for example,][]
{Griessmeier18Handbook,Lazio18Handbook,Zarka18Handbook}.

When setting up observation campaigns, the typical approach is to use 
the known exoplanetary parameters, and to estimate, planet by planet,
the maximum frequency of the emission (based on an estimate
\correction{of the planet's surface magnetic field strength})
 and the flux \correction{density of the radio emission} received at Earth. 
This basic approach \correction{has been used} 
at least since the seminal articles of 
\citet[][]{Zarka97} and \citet[][]{Farrell99}.
It immediately leads to \correction{several} criteria which make a planet a potentially
detectable source of radio emission: in order to have a maximum emission 
frequency above the terrestrial ionospheric cutoff, a certain magnetic 
moment is required; massive planets are more likely to have strong 
magnetic fields, which leads to better chances to detect their radio 
emission. 
Also, close-in planets are frequently assumed to be favorable for radio 
emission, as their proximity to the star leads to a larger energy
flux into the planetary magnetosphere. Assuming that a fixed 
\correction{fraction}
of that input power is converted \correction{to} radio power, close-in planets 
are usually among the favored targets.

Over the years, more accurate planetary parameters have been obtained,
approximations have been refined, some models have been added, 
and other models have been improved.
It is in this line of work that the previous 
implementation of the radio prediction code \citep{Griessmeier07AA}
is currently being replaced by the unified exoplanetary radio prediction 
code PALANTIR (Prediction Algorithm for star-pLANeT Interactions in Radio), 
which will be at the same time more user-friendly 
and easier to upgrade in the light of future theoretical work 
\citet{Mauduit23PRE9}. 

As our understanding of planetary radio emissions \correction{has grown}, and as the 
number of known planets \correction{has} literally exploded, the initially simple 
approximations have not only become \correction{more} complex, but also include more 
physical mechanisms.
%
%
One of the physical effects that has, so far, only been investigated
on a case-by-case basis, but which 
\correction{should be investigated}
\correction{in a more systematic way}, 
is the relation between the expected plasma conditions
in the vicinity of the planet and the detectibility of that planet via 
radio emission. 
\correction{Indeed, for low mass planets, 
strong stellar insolation can lead to an extended planetary 
atmosphere and thus a high local plasma density. If the plasma density, 
relative to the cyclotron frequency,
is too high, the radio emission is quenched.
For higher planetary masses, however, the planetary atmosphere is 
hydrostatic even for close-in orbits, and radio emission remains possible.}
\correction{This effect will be detailed in Section \ref{sec:plasma}.
Section \ref{sec:previous} will review previous studies. 
Based on these studies, Section \ref{sec:parameterspace} presents a  
parameter space where quenching can prevent radio emission from planets which 
might otherwise be considered good targets.
Section \ref{sec:perspectives} closes with concluding remarks 
and future perspectives.}


\section{Plasma conditions at the emission source site}
\label{sec:plasma}

Theoretical and observational work has shown that the CMI requires
specific plasma conditions to be able to operate. 
More precisely, for the CMI to operate, the
local electron plasma frequency  $f_{p}$ has to be smaller than the 
local electron cyclotron frequency $f_{c}$ by a certain factor. 
Theoretical work suggests a critical ratio of 0.4, i.e.~CMI emission 
can only operate if $f_{p}/f_{c} < 0.4$
\citep[e.g.][]{Lequeau85,Hilgers92,Zarka01cutoff}.
 
In other words, the CMI requires regions of low plasma density and 
strong magnetic field.
%
This condition is usually assumed to be fulfilled for exoplanets. 
For example, \citet{Griessmeier07AA} stated that 
the detectability of exoplanets are unlikely to be affected by 
high plasma density (including both stellar wind plasma and the planetary
environment).

However, for extremely close-in planets, specific conditions can arise.
Indeed, the upper atmosphere is heated by the high X-ray and 
extreme ultraviolet (EUV) -or XUV for X-ray and extreme ultraviolet-  
flux of the host star. In some cases, this can result in expanded upper
atmospheres. In this case, the ionized 
gas can prevent the generation and/or escape of radio emission. 
This effect has already been demonstrated by numerical 
simulations \citep{DaleyYates17,DaleyYates18}
as well as by analytical calculations 
\citep{Weber17pre,Weber17,Weber18mnras,Erkaev22}.

\section{Previous case studies}
\label{sec:previous}

In the literature, only a handful of cases have been studied so far:

\begin{itemize}
	\item	In the case of low mass Hot Jupiters 
			\correction{(with a mass similar 
			to that of Jupiter, or less)}, strong
			\correction{stellar} insolation (i.e. close orbital distance),
			can lead to an extended atmosphere, 
			and planetary radio emission \correction{is probably not} possible.
			\correction{\citet{Weber17pre,Weber17} show that this}
			is the case, for example, for the exoplanets
			HD 209458b \correction{(with a planetary mass of $0.69 \, M_J$)}
			and HD 189733b \correction{(with a planetary mass of $1.14 \, M_J$)}.
			For HD 209558b-like planets, 
			the critical orbital distance 
			\correction{(i.e.~the orbital distance below which this effect becomes important)}
			seems to be	somehere between 0.2 and 0.5 AU.
	\item	For high mass Hot Jupiters, the atmosphere is ``compact'', 
			i.e. strongly bound to the planet, 
			and radio emission is possible even for planets 
			at close orbital distances.
			This is the case, for example, for the exoplanet
			$\tau$ Bootis b (with an estimated planetary mass of 
			$5.84\, M_J$), as discussed by \citet{Weber18mnras}.
	\item	The case of $\upsilon$ Andromedae b is particularly interesting.
			The planet is known from \correction{radial velocity} observations.
			As a consequence, the true planetary mass is unknown. 
			Instead, the measurements only yield the projected mass, 
			which serves as a lower limit to the true planetary mass.
			The planetary mass being unknown, \citet{Erkaev22}
			have performed atmospheric models for different values of 
			the planetary mass. 
			They find that no radio emission should be visible if the 
			planetary mass $M$ is lower than $2.25 M_J$ (
			where $M_J$ is Jupiter's mass).
			On the other hand, \correction{planetary radio emission is 
			possible} if the planet is more massive, 
			i.e. if $M> 2.25 M_J$.
			This argument can also be turned around:  
			if radio emission is detected from this planet, this would be 
			a strong indication that the planetary mass is $>2.25 M_J$.
			It should be noted that both $\tau$ Bootis and $\upsilon$ Andromedae
			have been part of recent radio observations using LOFAR 
			(LOw Frequency ARray) and NenuFAR (New Extension in Nan\c{c}ay
			upgrading LOFAR)
			\citep{Turner21,Turner23PRE9}.
\end{itemize}

Clearly, the differences between planets are striking, and a coherent 
picture seems to start to emerge, which should briefly be compared to
the results of the typical approach (which ignores this atmospheric 
expansion effect).
In the simplified approach, a high planetary mass is favorable for radio 
detection, as it leads to a high maximum emission frequency (as mentioned above,
ground-based detection required at least $f_c^{min}=10$ MHz).
The simplified approach also favors close-in planets, where the power
input into the magnetosphere is high.

Taking into account the effect of potential quenching by a planetary 
expanded upper atmosphere, the picture slighty changes. 
A high planetary mass still is favorable for radio detection (as it 
decreases the ratio $f_{p}/f_{c}$). In addition, a high planetary mass
leads to a more strongly bound planetary atmosphere. Only for a low 
planetary mass the atmosphere can be extended, and lead to radio quenching.
On the other hand, small orbital distances are only favorable within limits:
if the orbital distance is below a critical value, the atmosphere becomes
highly heated and expands, leading to radio quenching.

\section{\correction{Parameter space for radio quenching}}
\label{sec:parameterspace}

\begin{figure}[!htb]
\centering
\includegraphics[width=1\textwidth]{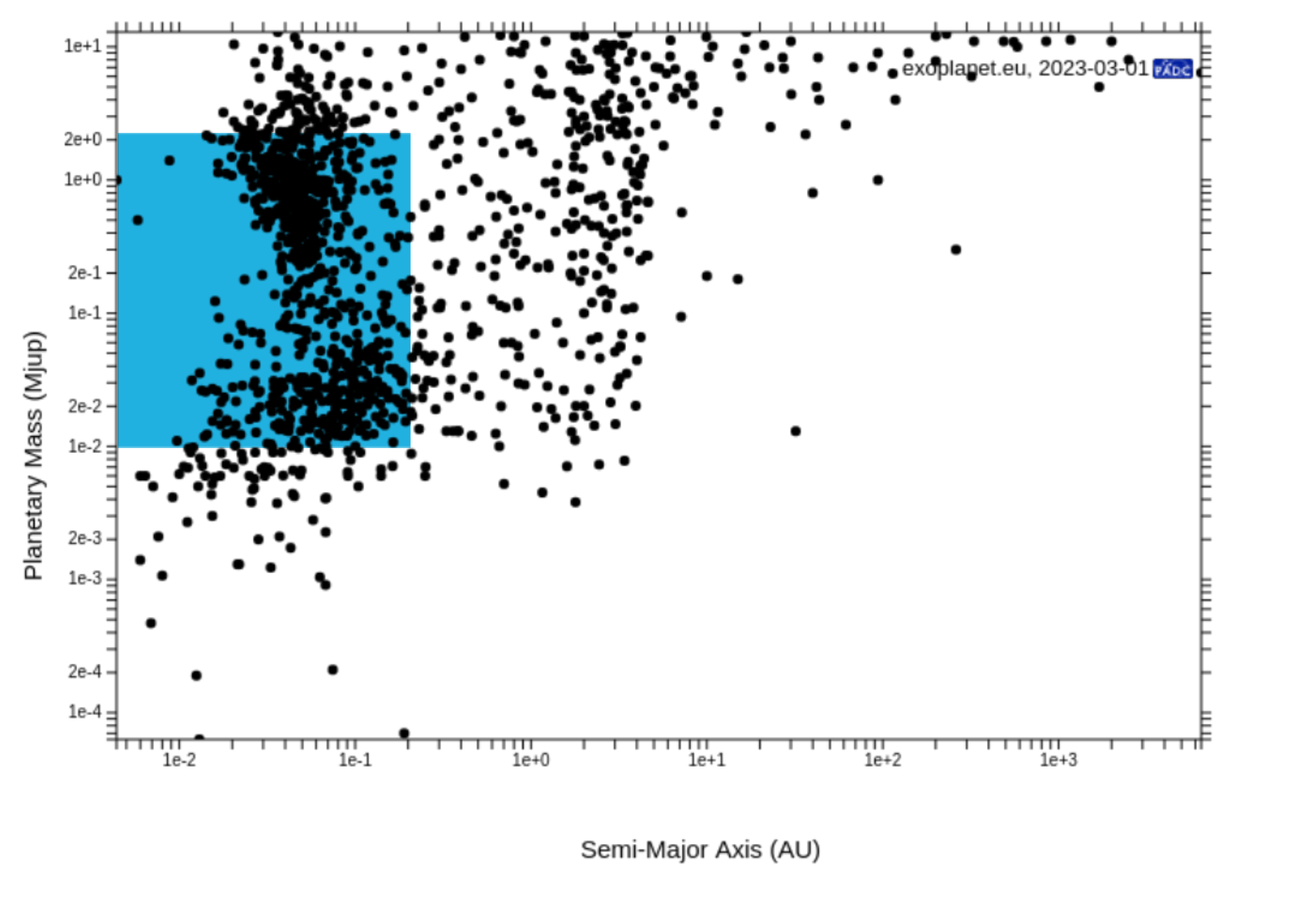}
\caption{Known explanets (as of 2023-03-01), based on \url{exoplanet.eu}
 \citep{2011AASchneider}. 
 Blue area: Planets with an orbital distance $d< 0.2$ AU, 
 and a planetary mass $M$ in the range $0.01<M\le2 M_J$). 
 In this parameter space, an extended planetary
 atmosphere \correction{could} potentially lead to radio quenching.
 }
 
\label{fig:stat}
\end{figure}

\correction{A detailed case-by-case calculation for each planet is 
beyond the scope of this work and is left for future work.
Here, we estimate for which region in the 
parameter space spanned by the planetary mass and orbital distance this
effect can become important, in the sense that 
it can prevent the generation of radio emission for 
otherwise good candidates.}
\correction{First, for the lower mass limit, we}
note that exoplanets with masses as low as 0.01 Jupiter masses
\correction{(and sometimes less)} 
are sometimes \correction{considered} as good candidates for the search for planetary radio
emission \citep{Mauduit23PRE9}. 
\correction{In principle, planets considerably less massive will be impacted 
even more strongly, but they are usually not considered as good candidates
for radio emission.
We thus adopt $M_\text{min}=0.01 M_J$ for our parameter space.}
\correction{Second, based on the case of $\upsilon$ Andromedae b 
(cf. Section \ref{sec:previous},
}
we assume that most planets with a 
mass \correction{above $\sim2 M_J$ are} protected against radio quenching,
and set $M_\text{max}=2 M_J$ for our parameter space.
\correction{Third, we
set a conservative lower limit of 0.2 AU for orbital distances 
where radio quenching may become important
(based on the case of HD209458 b, cf. Section \ref{sec:previous}).}
\correction{The region delimited by these three criteria 
is depicted in Figure \ref{fig:stat}).} 
%

\correction{Figure \ref{fig:stat}) shows that these criteria lead to}
a large parameter space where an extended atmosphere
may prevent planetary radio emission (see Figure \ref{fig:stat}). 
This parameter space currently includes 780 of the currently know 5332 exoplanets.
More importantly, many of the good candidates
suggested by radio scaling laws will fall into this parameter space.

\correction{In relaity, of course, both parameters are not independent, 
and the minimum planetary mass} required to prevent an extended 
planetary atmosphere will depend on the orbital distance, which will 
lead to a more complex shape than that shown in Figure \ref{fig:stat}.
The precise borders of the region in the parameter space spanned by planetary mass and 
orbital distance that is favorable for planetary radio emission still need to be determined. 
As additional parameters, this ``radio-favorable'' region will, also depend on
the planetary radius, the stellar mass and the stellar age.
This parameter space will be systematically explored in future work.

\section{Perspectives}
\label{sec:perspectives}

The importance of the planetary mass for the efficient generation of 
an intrinsic planetary magnetic field, and thus for the 
generation of planetary radio emission has long been realized. 
%
However, it becomes increasingly clear that a high planetary mass is 
important for a second reason: with a high mass, the planet can maintain
its evaporating atmosphere at close distance, and thus avoid conditions
where an extended ionized atmosphere can trap, or quench, the planetary
radio emission. 
The minimum required mass will depend on the planetary orbital distance
and the characteristics of its host star.

So far, this effect has only beeen studied for isolated cases.
We plan to perform a systematic study of the parameter space 
that is favorable for the generation and emission of planetary radio 
emission.
We also aim at including this criterion to the 
radio-prediction and target selection code PALANTIR
\correction{\citep{Mauduit23PRE9}}, which will allow to optimise the target selection
for observational campaigns with low frequency radio telescopes, 
such as the one currently ongoing at NenuFAR \citep{Turner23PRE9}.

\textit{Acknowledgements:} 

This work has made use of the Extrasolar Planet Encyclopaedia 
(exoplanet.eu) maintained by J. Schneider \citep{2011AASchneider}.

This work was supported by the Programme National de Plan\'{e}tologie 
(PNP) of CNRS/INSU co-funded by CNES 
and by the Programme National de Physique Stellaire (PNPS) of 
CNRS/INSU co-funded by CEA and CNES. 

\correction{We thank the anonymous referees for helpful and constructive suggestions.}

\newcommand{\newblock}{}
\bibliographystyle{mnras}
\bibliography{bibliography.bib}

\begin{thebibliography}{}
\makeatletter
\relax
\def\mn@urlcharsother{\let\do\@makeother \do\$\do\&\do\#\do\^\do\_\do\%\do\~}
\def\mn@doi{\begingroup\mn@urlcharsother \@ifnextchar [ {\mn@doi@}
  {\mn@doi@[]}}
\def\mn@doi@[#1]#2{\def\@tempa{#1}\ifx\@tempa\@empty \href
  {http://dx.doi.org/#2} {doi:#2}\else \href {http://dx.doi.org/#2} {#1}\fi
  \endgroup}
\def\mn@eprint#1#2{\mn@eprint@#1:#2::\@nil}
\def\mn@eprint@arXiv#1{\href {http://arxiv.org/abs/#1} {{\tt arXiv:#1}}}
\def\mn@eprint@dblp#1{\href {http://dblp.uni-trier.de/rec/bibtex/#1.xml}
  {dblp:#1}}
\def\mn@eprint@#1:#2:#3:#4\@nil{\def\@tempa {#1}\def\@tempb {#2}\def\@tempc
  {#3}\ifx \@tempc \@empty \let \@tempc \@tempb \let \@tempb \@tempa \fi \ifx
  \@tempb \@empty \def\@tempb {arXiv}\fi \@ifundefined
  {mn@eprint@\@tempb}{\@tempb:\@tempc}{\expandafter \expandafter \csname
  mn@eprint@\@tempb\endcsname \expandafter{\@tempc}}}

\bibitem[\protect\citeauthoryear{Daley-Yates \& Stevens}{Daley-Yates \&
  Stevens}{2017}]{DaleyYates17}
Daley-Yates S.,  Stevens I.~R.,  2017, Interacting fields and flows: {M}agnetic
  hot {J}upiters, \textit{Astron. Nachr.}, \textit{338, 881}

\bibitem[\protect\citeauthoryear{Daley-Yates \& Stevens}{Daley-Yates \&
  Stevens}{2018}]{DaleyYates18}
Daley-Yates S.,  Stevens I.~R.,  2018, Inhibition of the electron cyclotron
  maser instability in the dense magnetosphere of a hot jupiter,
  \textit{MNRAS}, \textit{479, 1194}

\bibitem[\protect\citeauthoryear{{Ergun}, {Carlson}, {McFadden}, {Delory},
  {Strangeway}  \& {Pritchett}}{{Ergun} et~al.}{2000}]{Ergun00}
{Ergun} R.~E.,  {Carlson} C.~W.,  {McFadden} J.~P.,  {Delory} G.~T.,
  {Strangeway} R.~J.,   {Pritchett} P.~L.,  2000, Electron-cyclotron {M}aser
  driven by charged-particle acceleration from magnetic field-aligned electric
  fields, \textit{ApJ}, \textit{538, 456}

\bibitem[\protect\citeauthoryear{Erkaev, Weber, Grie{\ss}meier, Lammer, Ivanov
  \& Odert}{Erkaev et~al.}{2022}]{Erkaev22}
Erkaev N.~V.,  Weber C.,  Grie{\ss}meier J.-M.,  Lammer H.,  Ivanov V.~A.,
  Odert P.,  2022, Can radio emission escape from the magnetosphere of
  $\upsilon$ andromedae b -- a new method to constrain the minimum mass of hot
  jupiters, \textit{MNRAS}, \textit{512, 4869}

\bibitem[\protect\citeauthoryear{Farrell, Desch  \& Zarka}{Farrell
  et~al.}{1999}]{Farrell99}
Farrell W.~M.,  Desch M.~D.,   Zarka P.,  1999, On the possibility of coherent
  cyclotron emission from extrasolar planets, \textit{J.~Geophys.~Res.},
  \textit{104, 14025}

\bibitem[\protect\citeauthoryear{Grie{\ss}meier}{Grie{\ss}meier}{2018}]{Griessmeier18Handbook}
Grie{\ss}meier J.-M.,  2018, in  eds Deeg H.~J.,  Juan Antonio~Belmonte J.~A.,
  , , Handbook of Exoplanets.
Springer, \mn@doi{10.1007/978-3-319-30648-3_159-1}

\bibitem[\protect\citeauthoryear{Grie{\ss}meier, Zarka  \&
  Spreeuw}{Grie{\ss}meier et~al.}{2007}]{Griessmeier07AA}
Grie{\ss}meier J.-M.,  Zarka P.,   Spreeuw H.,  2007, Predicting low-frequency
  radio fluxes of known extrasolar planets, \textit{A\&A}, \textit{475, 359}

\bibitem[\protect\citeauthoryear{Hilgers}{Hilgers}{1992}]{Hilgers92}
Hilgers A.,  1992, The auroral radiating plasma cavities, \textit{Geophys. Res.
  Lett.}, \textit{19, 237}

\bibitem[\protect\citeauthoryear{{Lazio}}{{Lazio}}{2018}]{Lazio18Handbook}
{Lazio} T. J.~W.,  2018, in  eds Deeg H.~J.,  Juan Antonio~Belmonte J.~A., , ,
  Handbook of Exoplanets.
Springer, \mn@doi{10.1007/978-3-319-55333-7_9}

\bibitem[\protect\citeauthoryear{{Le Qu\'{e}au}, Pellat  \& Roux}{{Le
  Qu\'{e}au} et~al.}{1985}]{Lequeau85}
{Le Qu\'{e}au} D.,  Pellat R.,   Roux A.,  1985, The maser synchrotron
  instability in an inhomogeneous medium : Application to the generation of the
  {A}uroral {K}ilometric {R}adiation, \textit{Ann. Geophys.}, \textit{3, 273}

\bibitem[\protect\citeauthoryear{{Mauduit}, {Grie{\ss}meier}  \&
  {Zarka}}{{Mauduit} et~al.}{2023}]{Mauduit23PRE9}
{Mauduit} E.,  {Grie{\ss}meier} J.-M.,   {Zarka} P.~{Turner} J.~D.,  2023,
  {PALANTIR: an updated prediction tool for exoplanetary radioemissions},
  \textit{in Planetary Radio Emissions IX,  eds Georg Fischer and Caitriona
  Jackman and Corentin Louis and Ali Sulaiman and Pietro Zucca}, Austrian
  Academy of Sciences Press, Vienna, pp xxx--xxx

\bibitem[\protect\citeauthoryear{{Schneider}, {Dedieu}, {Le Sidaner}, {Savalle}
   \& {Zolotukhin}}{{Schneider} et~al.}{2011}]{2011AASchneider}
{Schneider} J.,  {Dedieu} C.,  {Le Sidaner} P.,  {Savalle} R.,   {Zolotukhin}
  I.,  2011, {Defining and cataloging exoplanets: the exoplanet.eu database},
  \textit{\mn@doi [A\&A] {10.1051/0004-6361/201116713}}, \href
  {http://adsabs.harvard.edu/abs/2011A\%26A...532A..79S} {\textit{532, A79}}

\bibitem[\protect\citeauthoryear{Treumann}{Treumann}{2006}]{Treumann06}
Treumann R.~A.,  2006, The electron-cyclotron maser for astrophysical
  application, \textit{Astron. Astrophys. Rev.}, \textit{13, 229}

\bibitem[\protect\citeauthoryear{Turner et~al.,}{Turner
  et~al.}{2021}]{Turner21}
Turner J.~D.,  et~al., 2021, The search for radio emission from the
  exoplanetary systems 55 {C}ancri, $\upsilon$ {A}ndromedae, and $\tau$
  {B}o\"otis using {LOFAR} beam-formed observations, \textit{A\&A},
  \textit{645, A59}

\bibitem[\protect\citeauthoryear{{Turner}, {Zarka}, {Grie{\ss}meier}, {Mauduit}
   \& {Lamy}}{{Turner} et~al.}{2023}]{Turner23PRE9}
{Turner} J.~D.,  {Zarka} P.,  {Grie{\ss}meier} J.-M.,  {Mauduit} E.,   {Lamy}
  L.,  2023, {FOLLOW-UP RADIO OBSERVATIONS OF THE $\tau$ BO\"{O}TIS
  EXOPLANETARY SYSTEM: PRELIMINARY RESULTS FROM NENUFAR}, \textit{in Planetary
  Radio Emissions IX,  eds Georg Fischer and Caitriona Jackman and Corentin
  Louis and Ali Sulaiman and Pietro Zucca}, Austrian Academy of Sciences Press,
  Vienna, pp xxx--xxx

\bibitem[\protect\citeauthoryear{{Weber} et~al.,}{{Weber}
  et~al.}{2017a}]{Weber17pre}
{Weber} C.,  et~al., 2017a, {On the Cyclotron Maser Instability in
  Magnetospheres of Hot Jupiters - Influence of ionosphere models}, \textit{in
  Planetary Radio Emissions VIII, edited by G. Fischer, G. Mann, M. Panchenko,
  and P. Zarka, Austrian Academy of Sciences Press, Vienna, p.317-329, 2017},
  pp 317--329, \mn@doi{10.1553/PRE8s317}

\bibitem[\protect\citeauthoryear{{Weber} et~al.,}{{Weber}
  et~al.}{2017b}]{Weber17}
{Weber} C.,  et~al., 2017b, {How expanded ionospheres of Hot Jupiters can
  prevent escape of radio emission generated by the cyclotron maser
  instability}, \textit{\mn@doi [MNRAS] {10.1093/mnras/stx1099}}, \href
  {http://adsabs.harvard.edu/abs/2017MNRAS.469.3505W} {\textit{469, 3505}}

\bibitem[\protect\citeauthoryear{{Weber}, {Erkaev}, {Ivanov}, {Odert},
  {Grie{\ss}meier}, {Fossati}, {Lammer}  \& {Rucker}}{{Weber}
  et~al.}{2018}]{Weber18mnras}
{Weber} C.,  {Erkaev} N.~V.,  {Ivanov} V.~A.,  {Odert} P.,  {Grie{\ss}meier}
  J.-M.,  {Fossati} L.,  {Lammer} H.,   {Rucker} H.~O.,  2018, {Supermassive
  hot Jupiters provide more favourable conditions for the generation of radio
  emission via the cyclotron maser instability - a case study based on Tau
  Bootis b}, \textit{\mn@doi [MNRAS] {10.1093/mnras/sty2079}}, \href
  {http://adsabs.harvard.edu/abs/2018MNRAS.480.3680W} {\textit{480, 3680}}

\bibitem[\protect\citeauthoryear{Zarka}{Zarka}{1998}]{Zarka98}
Zarka P.,  1998, Auroral radio emissions at the outer planets: {O}bservations
  and theories, \textit{J.~Geophys.~Res.}, \textit{103, 20159}

\bibitem[\protect\citeauthoryear{{Zarka}}{{Zarka}}{2018}]{Zarka18Handbook}
{Zarka} P.,  2018, in  eds Deeg H.~J.,  Juan Antonio~Belmonte J.~A., , ,
  Handbook of Exoplanets.
Springer, \mn@doi{10.1007/978-3-319-55333-7_22}

\bibitem[\protect\citeauthoryear{Zarka et~al.,}{Zarka et~al.}{1997}]{Zarka97}
Zarka P.,  et~al., 1997, Ground-based high sensitivity radio astronomy at
  decameter wavelengths, \textit{in Planetary Radio Emissions IV,  eds H. O.
  Rucker and S. J. Bauer and A. Lecacheux}, Austrian Academy of Sciences Press,
  Vienna, pp 101--127

\bibitem[\protect\citeauthoryear{Zarka, Queinnec  \& Crary}{Zarka
  et~al.}{2001}]{Zarka01cutoff}
Zarka P.,  Queinnec J.,   Crary F.~J.,  2001, Low-frequency limit of {J}ovian
  radio emissions and implications on source locations and {I}o plasma wake,
  \textit{Planet. Space Sci.}, \textit{49, 1137}

\makeatother
\end{thebibliography}

\end{document}